\documentclass[conference]{IEEEtran}
\usepackage{multirow}
\usepackage{cite}

\ifCLASSINFOpdf
   \usepackage[pdftex]{graphicx}
   \usepackage{epsfig}
   \usepackage{epstopdf}
   \DeclareGraphicsExtensions{.pdf,.jpeg,.png, .eps}
\else
\fi
\usepackage[cmex10]{amsmath}
\usepackage{amsthm}
\usepackage{amssymb}
\usepackage{algorithm}
 \usepackage{algpseudocode} 

\usepackage{array}
\usepackage{fixltx2e}

\DeclareMathOperator*{\GTRLESS}{\gtrless}

\begin{document}
%
\title{Optimal Simultaneous Detection and Signal and Noise Power Estimation}

\author{\IEEEauthorblockN{Long Le, Douglas L. Jones}
\IEEEauthorblockA{Department of Electrical and Computer Engineering\\
University of Illinois at Urbana-Champaign}}
\maketitle

\begin{abstract}
Simultaneous detection and estimation is important in many engineering applications. In particular, there are many applications where it is important to 
perform signal detection and Signal-to-Noise-Ratio (SNR) estimation jointly. Application of existing frameworks in the literature that handle 
simultaneous detection and estimation is not straightforward for this class of application. This paper therefore aims at bridging the gap between an 
existing framework, specifically the work by Middleton et al., and the mentioned application class by presenting a jointly optimal detector and signal 
and noise power estimators. The detector and estimators are given for the Gaussian observation model with appropriate conjugate priors on the 
signal and noise power. Simulation results affirm the superior performance of the optimal solution compared to the separate detection and 
estimation approaches.
\end{abstract}


%
\IEEEpeerreviewmaketitle


\section{Introduction}
Traditional signal processing applications, such as radar, sonar and communication systems, are often limited to separate applications of
detection and estimation theory \cite{levy2008principles}. Parameters of interest are first estimated using pilot signals during the training period; 
then the estimates are fed into the detection process \cite{doug2004adapt}.

Many modern applications of detection and estimation theory, however, require the combined effort of both detector and estimator. A few examples
are
\begin{itemize}
\item In object search, tracking, and classification using mobile robots with limited resources, 
detection and estimation are the two main competing tasks on an energy-limited system. It is therefore important for the system to consider
both tasks jointly for the best utilization of the robot's  resource \cite{wang2011bayesian, wang2011risk}.
\item Recently,  it was shown in \cite{long2013energy} that a scheduler can rely on SNR estimates to 
elect between different detectors to yield an energy-efficient detection system. The efficiency of the detection system therefore depends on the quality of the 
SNR estimates. Incorporating the estimator design into the design of the detection system, instead of considering them separately, is evidently necessary 
for optimality.
\item In voice activity detector (VAD) designs, the speech detection performance depends heavily on the quality of 
the noise and {\it a priori} Signal-to-Noise-Ratio (SNR) estimates \cite{sohn1998voice, sohn1999statistical}; therefore, it is important for an optimal design 
of VAD to consider both the detection and estimation operations jointly.
\item Simultaneous detection and estimation can also be used to greatly improve the performance of existing techniques that treated the two operations 
separately. For example, Abramson and Cohen proposed a novel method for speech-enhancement that utilized simultaneous detection and estimation 
\cite{abramson2007simultaneous}. While traditional speech-enhancement systems that used spectral suppression 
\cite{ephraim1984speech, cappe1994elimination} only performed speech estimation and had to sacrifice either musical noise reduction or speech distortion 
in highly non-stationary noise environments, Abramson and Cohen's work allowed a systematic way to optimally trade-off between the musical noise 
reduction and speech distortion, hence significantly improving the performance of their speech-enhancement system.
\end{itemize}

Some of the above-mentioned applications can be solved readily using the existing literatures for optimal simultaneous detection and estimation (OSDE), 
such as the framework of Middleton et al. \cite{middleton1968simultaneous, fredriksen1972simultaneous} or
Moustakides et al.\cite{moustakides2011optimum, jajamovich2011optimal, moustakides2012joint}. However, for a certain class of applications
where the estimation of SNR, specifically signal and noise power, under signal presence uncertainty is required, it is unclear how the existing frameworks 
can be utilized (see Section \ref{sec:back}). The contribution of this work therefore aims at bridging the gap between the existing theoretical work, 
specifically by Middleton et al. \cite{middleton1968simultaneous}, and the referred application class by providing the joint optimal detector and 
estimators for signal and noise power.

The roadmap for the rest of the paper is given as follows. Section \ref{sec:back} gives an overview of the prior work in the field of simultaneous
detection and estimation. Section \ref{sec:form} formulates the problem based on the statistical decision framework of Middleton et al. \cite{middleton1968simultaneous}. The benefit of this formulation is then demonstrated in Section \ref{sec:soln} on the classical Gaussian observation 
model  with appropriate conjugate priors on the signal and noise power. The Gaussian model was chosen because it has been proven to be sufficient 
for many application classes, such as communications 
\cite{levy2008principles} and voice activity detection \cite{sohn1999statistical}. Finally, Section \ref{sec:sim} 
gives empirical evidence to validate the proposed method.

\section{Background}
\label{sec:back}

It is important to first distinguish between the {\it bona fide} simultaneous detection and estimation formulations, where the parameters to be estimated
are continuous, and the multiple-hypotheses 
formulations in \cite{hwang1992simultaneous, djuric1993simultaneous, baygun1995optimal}. In the multiple-hypotheses formulations, the latent 
parameters are discrete and finite. For instance, the parameters in \cite{hwang1992simultaneous, djuric1993simultaneous} are the frequency indices; 
the framework in \cite{baygun1995optimal} only considered parameters that live in discrete parameter spaces with finite elements. The finite parameter space reduces the estimation problem to the classification problem, where observations are classified
into one of the multiple hypotheses. Since detection is merely a special case of classification, the whole problem of "joint detection and classification" 
\cite{baygun1995optimal} is simply the classical multiple-hypotheses testing problem. 

There exist multiple frameworks in the literature to address the {\it bona fide} simultaneous detection and estimation problem.
The most widely used one is probably the frequentist framework due to its simple implementation. Specifically, the frequentist framework assumes no prior
knowledge on the distribution of the unknown parameters. Maximum-likelihood (ML) estimators are used to provide estimates for the parameters in
the likelihood-ratio test of the detector, yielding the celebrated generalized likelihood-ratio test (GLRT) \cite{levy2008principles, sayeed1995optimal}. 
Obviously, GLRT is suboptimal if prior distributions of parameters are known.

The Bayesian framework takes advantage of the prior distributions, of both the unknown hypotheses $H_0, H_1$ and the unknown parameters 
$\theta_0, \theta_1$, to minimize the Bayes risk of the joint detection and estimation operations. 
Middleton et al. \cite{middleton1968simultaneous, fredriksen1972simultaneous}
were the first to lay the foundation for this framework. Recently, Moustakides et al. \cite{moustakides2011optimum, jajamovich2011optimal} relaxed the 
prior distribution assumption on the hypotheses $H_0, H_1$ to propose a Neyman-Pearson-like formulation; the unknown parameters 
are still viewed as random with known distributions. 
In either the pure-Bayesian (Middleton) or the Neyman-Pearson-like formulation (Moustakides), 
the common observation model was 
given as follows.
\begin{equation}\label{eqn:obsModel}
\begin{aligned}
H_0: \mathbf{Y} \sim f_0(\mathbf{y} \vert \theta_0), \Theta_0 \sim p_0(\theta_0)\\
H_1: \mathbf{Y} \sim f_1(\mathbf{y} \vert \theta_1), \Theta_1 \sim p_1(\theta_1)
\end{aligned}
\end{equation}
where $\mathbf{Y}$ is the observation vector. Note that bold fonts are used to distinguish vector against scalar quantities.

While the theoretical observation model in \eqref{eqn:obsModel} is readily suitable for some problems 
\cite{moustakides2012joint, hwang1992simultaneous}, a more physically amenable observation model for many problems can be given as follows.
\begin{equation}\label{eqn:phyObsModel}
\begin{aligned}
H_0:& \mathbf{Y}  = \mathbf{N} \\
H_1:& \mathbf{Y} = \mathbf{N}  + \mathbf{X}
\end{aligned}
\end{equation}
where $\mathbf{X}$ is the signal vector with a scalar variance (power) $S$ and $\mathbf{N}$ is the noise vector with 
a scalar variance (power) $V$. In addition, $S$ and  $V$ are viewed as random parameters with known prior distribution, $p_S(s)$ and $p_V(v)$.
One way to put the observation model \eqref{eqn:phyObsModel} into the form of \eqref{eqn:obsModel} is to treat $\theta_1$ as a vector of
two components $s, v$ and $\theta_0$ as $v$.

In the next section, the problem of simultaneous detection and estimation of signal and noise power will be formulated and solved optimally. 
It is noteworthy to mention that this approach is different from 
most prior work that are heuristic-based. In particular, SNR estimation is commonly achieved by a noise tracker and an {\it a priori} SNR estimator, where 
each component is individually designed \cite{sohn1998voice, abramson2007simultaneous}. 
Techniques for noise tracking include soft decision for MMSE criterion \cite{sohn1998voice},
signal-presence-probability controlled recursive average \cite{cohen2002noise, cohen2003noise, gerkmann2012unbiased}, and minimum statistic \cite{martin2001noise}. Techniques for {\it a priori} SNR estimation include ML \cite{ephraim1984speech, sohn1998voice} and decision-directed (DD) \cite{ephraim1984speech, sohn1999statistical}.

\section{Formulation}
\label{sec:form}

This section formulates the simultaneous detection and SNR estimation problem based on the framework of Middleton et al. 
\cite{middleton1968simultaneous}. This allows the use of prior probabilities from both hypotheses and parameters in order to calculate the 
Bayes risk or expected cost associated with the combined detection and estimation operations.
\begin{equation}\label{eqn:risk}
R(\delta, \hat{s}, \hat{v}) = \mathbb{E}\big[ \sum^{1,1}_{\substack{i=0\\j=0}} C_{ij}(S,V, \hat{s}(\mathbf{Y}), \hat{v}(\mathbf{Y})) \pi_i \delta(\gamma_j|\mathbf{Y}))\big]
\end{equation}
where the cost functions $C_{ij}$ are assumed to have quadratic forms as follows. 
$$
\begin{aligned}
C_{11}(S, V, \hat{s}(\mathbf{Y}), \hat{v}(\mathbf{Y})) &= \big[(S - \hat{s}(\mathbf{Y}))^2 + (V - \hat{v}(\mathbf{Y}))^2\big]b_{11} \\&+ a_{11}\\
C_{01}(V, \hat{s}(\mathbf{Y}), \hat{v}(\mathbf{Y})) &= \big[(0 - \hat{s}(\mathbf{Y}))^2 + (V - \hat{v}(\mathbf{Y}))^2\big]b_{01} \\&+ a_{01}\\
C_{10}(S, V, \hat{v}(\mathbf{Y})) &= \big[(S - 0)^2 + (V - \hat{v}(\mathbf{Y}))^2\big]b_{10} \\&+ a_{10}\\
C_{00}(V,  \hat{v}(\mathbf{Y})) &= (V - \hat{v}(\mathbf{Y}))^2b_{00} + a_{00}
\end{aligned}
$$
with the signal power estimate $\hat{s}(\mathbf{Y})$ being 0 when the detector decides a noise-only vector; the noise power estimate 
$\hat{v}(\mathbf{Y})$ is always provided. The prior probabilities of each hypothesis are denoted by 
$\pi_0 = P(H_0)$ and $\pi_1 = P(H_1)$. $\delta$ is the decision rule that governs the distribution of the decisions random
variable that takes on value $\gamma_1, \gamma_0$ given the observation vector $\mathbf{Y}$. Finally, for a decision $\gamma_j$ and a true hypothesis $H_i$, $a_{ij}$  is the detection cost parameter;
$b_{ij}$ is the conversion parameter that maps estimation error into detection error, hence specifying the trade-off between detection and 
estimation cost. The choice of these parameters directly affects the resulting joint detector and estimator, as will be seen later in this section.

It is worth mentioning that the conversion parameter $b_{ij}$ should be chosen to take into account the scaling of data. Scaled data leads
to scaled estimation error, which needs readjustment in accordance with the detection error.

Define the following conditional risks, similarly to what was done in \cite{abramson2007simultaneous}
\begin{equation}\label{eqn:condRisks}
\begin{aligned}
r_{11}(\mathbf{y}, \hat{s}, \hat{v}) &= \int \int C_{11}(s, v, \hat{s}(\mathbf{y}), \hat{v}(\mathbf{y})) f_1(\mathbf{y}| s, v) p_{S,V}(s,v) \\
ds\ dv\\
r_{01}(\mathbf{y}, \hat{s}, \hat{v}) &= \int C_{01}(v, \hat{s}(\mathbf{y}), \hat{v}(\mathbf{y})) f_0(\mathbf{y}| v) p_V(v) dv\\
r_{10}(\mathbf{y}, \hat{v}) &= \int \int C_{10}(s, v, \hat{v}(\mathbf{y})) f_1(\mathbf{y}| s, v) p_{S,V}(s, v) ds\ dv\\
r_{00}(\mathbf{y}, \hat{v}) &= \int C_{00}(v, \hat{v}(\mathbf{y})) f_0(\mathbf{y}| v) p_V(v) dv
\end{aligned}
\end{equation}
where $f_1(\mathbf{y}|s,v)$ and $f_0(\mathbf{y}|v)$ are observation distributions under $H_1$ and $H_0$, respectively.
In general, the signal and noise power can be statistically dependent as denoted by $p_{S,V}(s,v)$.
Hence \eqref{eqn:risk} can be rewritten explicitly as follows
$$
\begin{aligned}
R(\delta, \hat{s}, \hat{v}) = &\int \delta(\gamma_1 | \mathbf{y}) \big[\pi_1r_{11}(\mathbf{y}, \hat{s}, \hat{v}) + \pi_0 r_{01}(\mathbf{y}, \hat{v})\big]dy +\\
&\int \delta(\gamma_0 | \mathbf{y}) \big[ \pi_1r_{10}(\mathbf{y}, \hat{v}) + \pi_0r_{00}(\mathbf{y}, \hat{s}, \hat{v})\big] dy
\end{aligned}
$$
and the objective is to minimize it with respect to the decision rule and the estimators. Namely,
$$
\min_{\delta, \hat{s}, \hat{v}} R(\delta, \hat{s}, \hat{v})
$$

The solution for the above minimization problem is straightforward \footnote{Using the two-step minimization procedure 
from \cite{middleton1968simultaneous}} and intuitive, therefore only results are given while derivations are left out
due to page limitation. For simplicity, the following shorthand notations are used.
$$
\begin{aligned}
&\langle f(s, v) \rangle_{S,V} = \int \int f(s, v) p_{S,V}(s,v) ds dv \\
&\langle f(v) \rangle_{V} = \int f(v) p_V(v) dv
\end{aligned}
$$
for any function $f$ such that the integral is well-defined.

The optimal signal-power estimate is given by the following expression.
\begin{equation}\label{eqn:optSigPowEst}
\hat{s}^{opt} = \frac{\Lambda_1}{\Lambda_1+1} \hat{s}^{H_1}
\end{equation}
where 
$$
\Lambda_1 = \frac{b_{11}\pi_1\langle  f_1(\mathbf{y}|s, v) \rangle _{S,V}}{b_{01}\pi_0 \langle f_0(\mathbf{y}|v)\rangle_V}
$$
is the generalized likelihood ratio \cite{middleton1968simultaneous} when the detector's decision is $\gamma_1$ and
\begin{equation}\label{eqn:rawsig}
\hat{s}^{H_1} = \frac{\langle  s f_1(\mathbf{y}|s, v)\rangle_{S, V}}{\langle  f_1(\mathbf{y}|s, v)\rangle_{S, V}}
\end{equation}
is the signal power estimate assuming that $H_1$ is true. It is interesting to note that Equation \eqref{eqn:optSigPowEst}
has an intuitive interpretation: The optimal signal power estimator is simply the estimator assuming $H_1$ is true weighted by the 
posterior probability that $H_1$ is true, i.e. $\frac{\Lambda_1}{\Lambda_1 + 1}$. 

Unlike the single-equation signal-power estimate in \eqref{eqn:optSigPowEst}, the optimal noise-power estimate is given by the 
following two equations, depending on the decision of the detector.
\begin{equation}\label{eqn:optNoiPowEst1}
\hat{v}^{opt}_{\gamma_1} = \frac{\Lambda_1}{\Lambda_1 + 1} \hat{v}^{H_1} + \frac{1}{\Lambda_1+1} \hat{v}^{H_0}
\end{equation}
\begin{equation}\label{eqn:optNoiPowEst0}
\hat{v}^{opt}_{\gamma_0} = \frac{\Lambda_0}{\Lambda_0 + 1} \hat{v}^{H_1} + \frac{1}{\Lambda_0+1} \hat{v}^{H_0}
\end{equation}
where
$$
\Lambda_0 = \frac{b_{10}\pi_1 \langle f_1(\mathbf{y}|s, v) \rangle _{S,V}}{b_{00}\pi_0 \langle f_0(\mathbf{y}|v)\rangle_V}
$$
is the generalized likelihood ratio \cite{middleton1968simultaneous} when the detector's decision is $\gamma_0$ and
\begin{equation}\label{eqn:rawnoi}
\begin{aligned}
\hat{v}^{H_1} &= \frac{ \langle v  f_1(\mathbf{y}|s, v)  \rangle_{S, V}}{\langle  f_1(\mathbf{y}| s, v) 
 \rangle_{S, V}}\\
\hat{v}^{H_0} &= \frac{ \langle v  f_0(\mathbf{y}|v)\rangle_{V}}{\langle f_0(\mathbf{y}| v)\rangle_{V}}
\end{aligned}
\end{equation}
are the noise power estimates assuming either $H_1$ or $H_0$ was true, respectively. Similar to the optimal signal-power estimator, the optimal 
noise-power estimators in \eqref{eqn:optNoiPowEst1} and \eqref{eqn:optNoiPowEst0} also have intuitive interpretations: they are 
the weighted sum of the noise-power estimators under $H_1$ and $H_0$, with the weighting coefficients being
the posterior probabilities. \eqref{eqn:optNoiPowEst1} is used when the detector's decision is $\gamma_1$
while  \eqref{eqn:optNoiPowEst0} is used when the detector's decision is $\gamma_0$.

Based on the optimal signal power estimator in \eqref{eqn:optSigPowEst} and noise power estimators in \eqref{eqn:optNoiPowEst1}
and \eqref{eqn:optNoiPowEst0}, it can be shown that the optimal detector is
$$
\delta^{opt}(\gamma_1|\mathbf{y}) = 
\begin{cases}
1 &\text{ if } \frac{r_{10}(\mathbf{y}, \hat{v}^{opt}_{\gamma_0}) - r_{11}(\mathbf{y}, \hat{s}^{opt}, \hat{v}^{opt}_{\gamma_1})}
{r_{01}(\mathbf{y}, \hat{s}^{opt},\hat{v}^{opt}_{\gamma_1}) - r_{00}(\mathbf{y}, \hat{v}^{opt}_{\gamma_0})} \geq \frac{\pi_0}{\pi_1}\\
0 &\text{ else } 
\end{cases}
$$
Furthermore, if $b_{10} = a_{10} =b_{11}$, $a_{11} = 0$ and $b_{01} = a_{01} =b_{00}$, $a_{00} = 0$, then
the right hand side of \eqref{eqn:detector} can be compactly expressed as follows.
\begin{equation}\label{eqn:detector}
\frac{\langle f_1(\mathbf{y}|s, v) \rangle_{S,V}}{\langle f_0(\mathbf{y}|v) \rangle_{V}} 
\Big\lbrack 1-\hat{s}^{opt} + \frac{2\hat{s}^{opt}\hat{s}^{H_1}}{(\hat{s}^{opt}+1)}\Big\rbrack
\frac{b_{11}}{b_{00}}\gtrless \frac{\pi_0}{\pi_1}
\end{equation}
which is still fundamentally a generalized likelihood ratio. The extra weighting term exists to account for a joint detection and estimation operation. 
As a sanity check, if no estimation is required, i.e. $\hat{s}^{opt} = 0$, the detection rule in \eqref{eqn:detector} degenerates into 
the well-known generalized likelihood ratio test.


\section {Gaussian observations with appropriate conjugate priors on signal and noise power}
\label{sec:soln}
Expressions \eqref{eqn:optSigPowEst}, \eqref{eqn:optNoiPowEst1}, \eqref{eqn:optNoiPowEst0}, and \eqref{eqn:detector} 
all involve integrating some functions of the likelihoods multiplied by priors on the signal and noise power. Therefore with appropriate conjugate priors,
the analytical expressions for \eqref{eqn:optSigPowEst}, \eqref{eqn:optNoiPowEst1}, \eqref{eqn:optNoiPowEst0}, and \eqref{eqn:detector}
can be obtained. In particular, if the observation vectors are
i.i.d., zero-mean Gaussian, and the signal and noise are assumed to be independent under $H_1$, i.e.
$$
\begin{aligned}
H_0:& \mathbf{Y} \sim \frac{1}{\sqrt{2\pi v}^N} \exp{\frac{-\sum_{i=1}^N y_i^2}{2v}}\\
H_1:& \mathbf{Y} \sim \frac{1}{\sqrt{2\pi (s+v)}^N} \exp{\frac{-\sum_{i=1}^N y_i^2}{2(s+v)}}
\end{aligned}
$$

Under $H_0$, it is well-known that the conjugate prior for 
a Gaussian distribution with random variance is the inverse-Gamma distribution
\cite{gelman2003bayesian, murphy2007conjugate}. Hence, it is assumed that 
$$
p_V(v)  = \frac{\beta_0^{\alpha_0}}{\Gamma(\alpha_0)} \frac{1}{v^{\alpha_0+1}} \exp{\Big(-\frac{\beta_0}{v}\Big)}
$$
where $\alpha_0 > 0$ and $\beta_0 > 0$  are the shape and scale parameters under $H_0$. On the other hand, a natural conjugate prior under $H_1$ 
can be found to be
\begin{equation}\label{eqn:conjPrior}
p_{S,V}(s,v)  = \frac{\beta_1^{\alpha_1-1}}{C_{11}\Gamma(\alpha_1-1)} \frac{1}{(s+v)^{\alpha_1+1}} \exp{\Big(-\frac{\beta_1}{s+v}\Big)}
\end{equation}
where $\alpha_1 > 1$ and $\beta_1 > 0$  are the shape and scale parameters under $H_1$ and 
\begin{equation}\label{eqn:C}
C_{mn} = \int_{\phi_1}^{\phi_2} 2 \vert \sin^m\theta\cos^n\theta \vert d\theta \quad m,n \in \mathbb{N}^+
\end{equation}
is the normalizing constant that, in the case of $C_{11}$, ensures \eqref{eqn:conjPrior} is a distribution; 
hence it depends on the support of \eqref{eqn:conjPrior}, which is application-dependent.  
$\phi_1$ and $\phi_2$ are additional degrees of freedom that, once given, can be used to compute $C_{mn}$.
The proof that \eqref{eqn:conjPrior} is indeed a distribution follows from a straightforward change of variables.

The same change of variables also leads to the following results. Under $H_0$,
$$
\begin{aligned}
\langle f_0(\mathbf{y}|v) \rangle_{V} =& \frac{\beta_0^{\alpha_0}}{\sqrt{2\pi}^N\Gamma(\alpha_0)} 
\frac{\Gamma(\alpha_0+N/2)}{\large(\frac{2\beta_0+\sum_{i=1}^N y_i^2}{2}\large)^{\alpha_0+N/2}}\\
\langle v f_0(\mathbf{y}| v) \rangle_{V} =&  \frac{\beta_0^{\alpha_0}}{\sqrt{2\pi}^N\Gamma(\alpha_0)} 
\frac{\Gamma(\alpha_0+N/2-1)}{\large(\frac{2\beta_0+\sum_{i=1}^N y_i^2}{2}\large)^{\alpha_0+N/2-1}}
\end{aligned}
$$
and under $H_1$,
$$
\begin{aligned}
\langle f_1(\mathbf{y}|s, v) \rangle_{S,V} =& \frac{\beta_1^{\alpha_1-1}}{\sqrt{2\pi}^N\Gamma(\alpha_1-1)} 
\frac{\Gamma(\alpha_1+N/2-1)}{\large(\frac{2\beta_1+\sum_{i=1}^N y_i^2}{2}\large)^{\alpha_1+N/2-1}} \\
\langle v f_1(\mathbf{y}|s, v) \rangle_{S,V} =& \frac{C_{31}/C_{11}\beta_1^{\alpha_1-1}}{\sqrt{2\pi}^N\Gamma(\alpha_1-1)} 
\frac{\Gamma(\alpha_1+N/2-2)}{\large(\frac{2\beta_1+\sum_{i=1}^N y_i^2}{2}\large)^{\alpha_1+N/2-2}} \\
\langle s f_1(\mathbf{y}|s, v) \rangle_{S,V} =& \frac{C_{13}/C_{11} \beta_1^{\alpha_1-1}}{\sqrt{2\pi}^N\Gamma(\alpha_1-1)} 
\frac{\Gamma(\alpha_1+N/2-2)}{\large(\frac{2\beta_1+\sum_{i=1}^N y_i^2}{2}\large)^{\alpha_1+N/2-2}} \\
\end{aligned}
$$
These expressions are handy for computing \eqref{eqn:optSigPowEst}, \eqref{eqn:optNoiPowEst1}, \eqref{eqn:optNoiPowEst0},
and \eqref{eqn:detector}. 

\section{Simulation}
\label{sec:sim}

\begin{figure}[t]
\centering
\includegraphics[width=\linewidth]{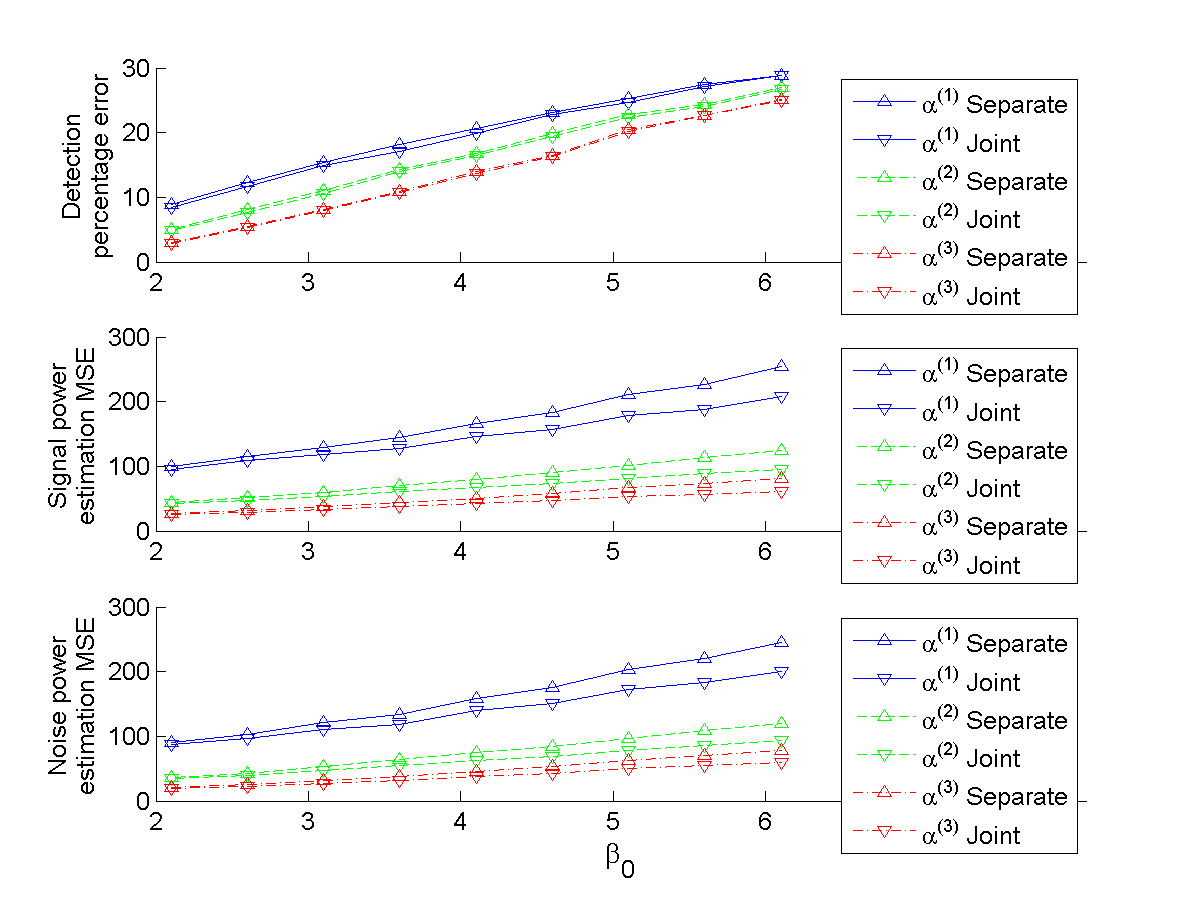}
\caption{Performance comparison of detectors and  estimators designed by the two approaches. In these simulations, the shape parameters
are assumed to be the same $\alpha_0 = \alpha_1 = \alpha$, and the signal power scale is fixed $\beta_1 = 9.1$  while the noise power scale varies.}
\label{fig:osdePlot}
\end{figure}

To demonstrate the utility of the derived detector and estimators, a set of simulations was carried out. For each simulation, 
20,000 independent observation vectors, each of size 128, were randomly generated using a zero-mean Gaussian distribution with 
hypothesis-dependent power. The hypothesis $H_1$ and $H_0$ are equally likely and the power under $H_0$ and $H_1$ are generated 
using the inverse gamma distribution.
The generic cost constants were set as follows  $b_{00} = b_{11} = b_{01} = b_{10} = a_{01} = a_{10} = 1$ and $a_{00} = a_{11} = 0$
for simplicity. Finally, $\phi_1$ is set to 0 and $\phi_2$ to $\pi/8$ to impose the prior knowledge that signal power is usually much higher 
than noise power.

The proposed simultaneous approach was compared against the separate approach. The separate detection
and estimation approach optimizes the detector and estimator separately. In summary,
the detector's test statistic is given by the generalized likelihood ratio
$$
\Lambda = \frac{ \langle f_1(\mathbf{y}|s, v) \rangle _{S,V}}{ \langle f_0(\mathbf{y}|v)\rangle_V}  \GTRLESS^{H_1}_{H_0} \frac{\pi_0}{\pi_1}
$$
and the estimators are simply $\hat{s}^{H_1}$, $\hat{v}^{H_1}$, and $\hat{v}^{H_0}$.

Figure \ref{fig:osdePlot} shows that, for all three criteria, the optimal joint design approach outperforms the separate approach.

\section{Conclusion}
\label{sec:concl}

An optimal detector and signal and noise power estimators was jointly derived for the Gaussian observation model with appropriate conjugate priors on 
the signal and noise power. Future work will apply the developed techniques to improve widely used algorithms, such as Sohn's VAD \cite{sohn1998voice}.


\section*{Acknowledgements}
This work was supported in part by TerraSwarm, one of six centers of STARnet, a Semiconductor Research Corporation program sponsored by MARCO and DARPA.



\bibliographystyle{IEEEtran}
\bibliography{osde}
%
%
%

\end{document}